\documentclass[preprint2]{aastex}
\usepackage{graphicx}
\usepackage{amssymb}
\usepackage{amsmath}
\usepackage{sidecap}
\shorttitle{Magnetic fluctuations at proton scales}
\shortauthors{Bruno and Telloni}
\begin{document}
\title{Spectral Features of Magnetic Fluctuations\\
at Proton Scales from Fast to Slow Solar Wind}
\author{R. Bruno\altaffilmark{1} and D. Telloni\altaffilmark{2}}

\altaffiltext{1}{National Institute for Astrophysics, Institute for Space Astrophysics and Planetology, Via del Fosso del Cavaliere 100, 00133 Roma, Italy}
\altaffiltext{2}{National Institute for Astrophysics, Astrophysical Observatory of Torino, Via Osservatorio 20, 10025 Pino Torinese, Italy}
\begin{abstract}
This Letter investigates the spectral characteristics of the interplanetary magnetic field fluctuations at proton scales during several time intervals chosen along the speed profile of a fast stream. The character of the fluctuations within the first frequency decade, beyond the high frequency break
located between
the fluid and the kinetic regime, strongly depends on the type of wind. While the fast wind shows a clear signature of both right handed and left handed polarized fluctuations, possibly associated with KAW and Ion-Cyclotron waves, respectively, the rarefaction region, where the wind speed and the Alfv\'{e}nicity of low frequency fluctuations decrease, shows a rapid disappearance of the ion-cyclotron signature followed by a more gradual disappearance of the KAWs. Moreover, also the power associated to perpendicular and parallel fluctuations experiences a rapid depletion, keeping, however, the power anisotropy in favour of the perpendicular spectrum.

\end{abstract}
\section{Introduction}
\label{sec:introduction}

Interplanetary magnetic field fluctuations show a clear turbulent spectrum characterized by a well established Kolmogorov scaling \citep[see the reviews by][and references therein]{tumarsch1995, brunocarbone2013}.
Energy cascades from the largest energy-containing eddies to the high frequency region of the spectrum where wave-particle interactions energize the ions at scales comparable with the typical proton scales \citep{marsch2012}.

The temperature anisotropy shown by the proton velocity distribution,
as well as the preferential heating and acceleration of minor ions \citep[see the review by][]{marsch2006} are clear indicators of the coupling between the magnetic energy of the fluctuations and the kinetic energy of the ions. The consequent effect of this energy transfer is a steepening of the spectral index, which marks the beginning of the kinetic range \citep{denskat1983} although, a steepening of the spectrum at proton scales can also be obtained without invoking dissipation, taking into account the Hall effect \citep{galtier2007}. The location of this high frequency break depends on the local magnetic field and plasma conditions and
varies with the heliocentric distance, moving to lower frequencies as the wind expands, as shown by \citet{bruno2014a}. These authors concluded that an ion-cyclotron resonance dissipation mechanism, in which the Doppler-shifted frequency matches the particle gyrofrequency, must participate in the spectral cascade together with other possible kinetic non-resonant mechanisms, as the Landau damping acting on the perpendicular short-scale fluctuations generated by the large-scale eddies.

On the other hand, the nature of the fluctuations within the kinetic range is still debated \citep{alexandrova2013}. The fact that one of the properties of the transition range separating the fluid from the kinetic regime is represented by an increase of compressibility \citep{alexandrova2008} suggested the presence of Kinetic Alfv$\acute{\textrm{e}}$n Waves \citep[KAWs, hereafter,][]{leamon1998,alexandrova2008,hamilton2008,sahraoui2009,turner2011,kiyani2013} and/or whistler waves \citep{gary2004,gary2009,tenbarge2012}.
In this respect, \citet{salem2012}, using Cluster observations in the solar wind, showed that the properties of the small-scale fluctuations are inconsistent with the whistler wave model, but strongly agree with the prediction of a spectrum of KAWs with nearly perpendicular wavevectors.


Moreover, \cite{alexandrova2008} reported that the intermittency character of magnetic field fluctuations within the kinetic range increases towards smaller scales and  persists at least to electron scales \citep{perri2012,wan2012,karimabadi2013}, indicating the presence of coherent magnetic structures advected by the solar wind. Further analyses associated elevated plasma temperature and anisotropy events with these structures suggesting that inhomogeneous dissipation was at work \citep{servidio2012}. Partially at odds with these results, \citet{wu2013}, using both flux-gate and search-coil magnetometers onboard Cluster, found kinetic scales much less intermittent than fluid scales. These authors recorded a remarkable and sudden decrease back to near-Gaussian values of intermittency around scales of about ten times the ion inertial scale, followed by a modest increase moving towards electron scales, in agreement with \citet{kiyani2009}, who showed observational results suggesting a scale-invariance within the small-scale range.


Other authors \citep{he2011,he2012a,he2012b,podesta2011} studied the polarization state of the fluctuations, within the kinetic regime, adopting a wavelet transform of the reduced magnetic helicity \citep{matthaeus1982,bruno2008} observed in a plane perpendicular to the sampling direction.
This kind of analysis was done for different pitch angles $\theta_{VB}$ between the the flow direction and the local mean magnetic field \citep{horbury2008}.


These authors found left-handed Alfv$\acute{\textrm{e}}$n/ion-cyclotron waves propagating outward almost parallel to the local magnetic field and right-handed KAWs propagating at large angles which confirmed previous conclusions by \citet{goldstein1994,leamon1998,hamilton2008} about the presence of KAWs.


In a recent study \citet{telloni2015} found that the spectral location of these two populations follow the frequency shift experienced by the high frequency spectral break during the radial expansion of the wind \citep{bruno2014a}. This behavior was interpreted as a further experimental evidence relating the presence of these fluctuations to the location of the frequency break. The same authors suggested that the decrease of intermittency found beyond the spectral break might be an effect of the stochastic nature of these fluctuations. Finally, \citet{bruno2014b} showed that the spectral slope is generally higher whenever the power level (and/or the Alfv\'{e}nic character) of the fluctuations is higher within the inertial range. They suggested that the behavior of the spectral slope might be related to some dissipative mechanism, like Landau damping and/or ion-cyclotron resonance.

Consequently, it is interesting to verify whether the polarization of the fluctuations
changes when we move from fast to slow wind within the same high speed stream given that fluctuations in the inertial range are progressively characterized by different Alfv\'{e}nicity, compressibility and intermittency \citep{brunocarbone2013}. This is the main goal of the present study.

\section{Data analysis and results}
\label{sec:data_analysis}
We adopt the same data analysis techniques reported in \cite{telloni2015} to study the polarization character of the fluctuations at kinetic scales as a function of the pitch angle between the sampling direction and the local magnetic field direction. The study will be performed scale by scale, sampling different field and plasma intervals along the wind speed profile, from fast to slow wind.
We chose a fast wind stream observed by WIND between the end of June and the beginning of July 2010, the same stream that was studied in \cite{telloni2015}. High resolution magnetic field measurements at about $92 \textrm{ms}$, were taken by the Magnetic Field Instrument \citep[MFI,][]{lepping1995} onboard Wind while, $1\textrm{min}$ plasma measurements were performed by the Solar Wind Experiment \citep[SWE,][]{ogilvie1995}.
Both datasets are available at the NASA-CDAWEB facility.

\begin{figure*}
	\centering
	\includegraphics[scale=0.6]{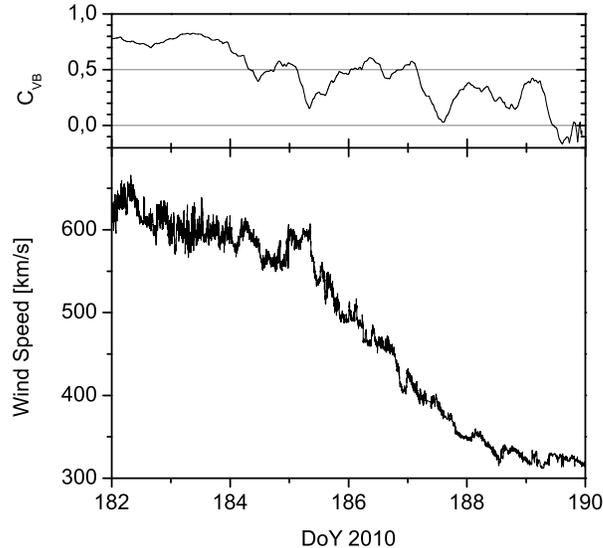}
	\caption{top panel: Time profile of the correlation factor $C_{VB}$ at the scale of $1\,h$ and smoothed out with a 12-hour sliding window; bottom panel: wind speed profile at time resolution of 1 min.}
	\label{fig:correlation_factor}
\end{figure*}

The speed profile of this fast wind stream is characterized by large amplitude velocity fluctuations within the trailing edge and much smaller fluctuations within the following rarefaction region \citep{hundhausen1972}, where the speed decreases faster. These strong fluctuations are intimately related to the presence of large amplitude Alfv\'{e}nic fluctuations \citep{matteini2014} as can be inferred from Figure \ref{fig:correlation_factor} where we show the speed profile together with the value of the correlation coefficient $C_{VB}$ \citep{bavassano1998}.
This parameter estimates the correlation level between magnetic field and velocity fluctuations being expressed like $C_{VB}={\sigma_{c}}/\sqrt{1-\sigma_{r}^{2}}$ where, $\sigma_{c}$ and $\sigma_{r}$ are the normalized forms of  cross-helicity and residual energy, respectively \citep{brunocarbone2013}, and $\sigma_{r}\neq\pm 1$. The parameter $C_{VB}$ has been evaluated on a time scale of $1 \textrm{hr}$ and successively, for graphical reasons, results have been smoothed with a $12 \textrm{hr}$ sliding average. Thus, $C_{VB}$ shows the level of the Alfv\'{e}nic correlations along the speed profile of this stream. Before the beginning of DoY 184, $C_{VB}$ is largely above 0.5 in magnitude. Afterwards, this parameter experiences a fast decrease towards 0, especially within the stream rarefaction region. The positive sign of the correlation indicates that the Alfv\'{e}nic fluctuations have an outward sense of propagation since the background magnetic field is inward directed.

We recall briefly the definitions and the technique we used, suggesting to refer to \citet{telloni2015} for a more detailed discussion.
The polarization of the fluctuations can be studied using the normalized reduced magnetic helicity $\sigma_{m}$ \citep{matthaeus1982}.
This measurable can be investigated in both time $t$ and temporal scale $\tau$ by means of the wavelet transforms \citep{torrence1998}, as suggested by \citet{bruno2008} and successively adopted by \citet{he2011}, \citet{he2012a}, \citet{he2012b}, \citet{podesta2011}.




The normalized reduced magnetic helicity $\sigma_{m}(t,\tau)$ varies between $+1$ and $-1$, positive and negative sign for left and right circular polarization, respectively. For an inward oriented background magnetic field, assuming outward propagation, a left-handed ion-cyclotron waves would have a positive magnetic helicity. The same wave would result with a negative helicity for an outward oriented magnetic field \citep{narita2009,he2011}.
Thus, it is necessary to know the angle $\theta_{VB}$ between the sampling direction, assumed along the wind direction, and the magnetic field, scale by scale. In order to do so, we first reorder all the values of $\sigma_{m}(t,\tau)$ into values of $\sigma_{m}(\theta_{VB},\tau)$. Then, to determine the local magnetic field, scale by scale, we operate a convolution between a Gaussian (normalized to unity), whose width is equal to the scale $\tau$, and the magnetic field $\textbf{B}_{0}(t)$ \citep{horbury2008,he2011,podesta2011}.
Finally, all the values of $\sigma_{m}(t,\tau)$, found within the same angular bin of $\theta_{VB}$, are averaged together in order to obtain the distribution of $\sigma_{m}(\theta_{VB},\tau)$. In our case, the angular step of this distribution is $1^{\circ}$ wide.





\begin{SCfigure*}
	\centering
\includegraphics[scale=0.70]{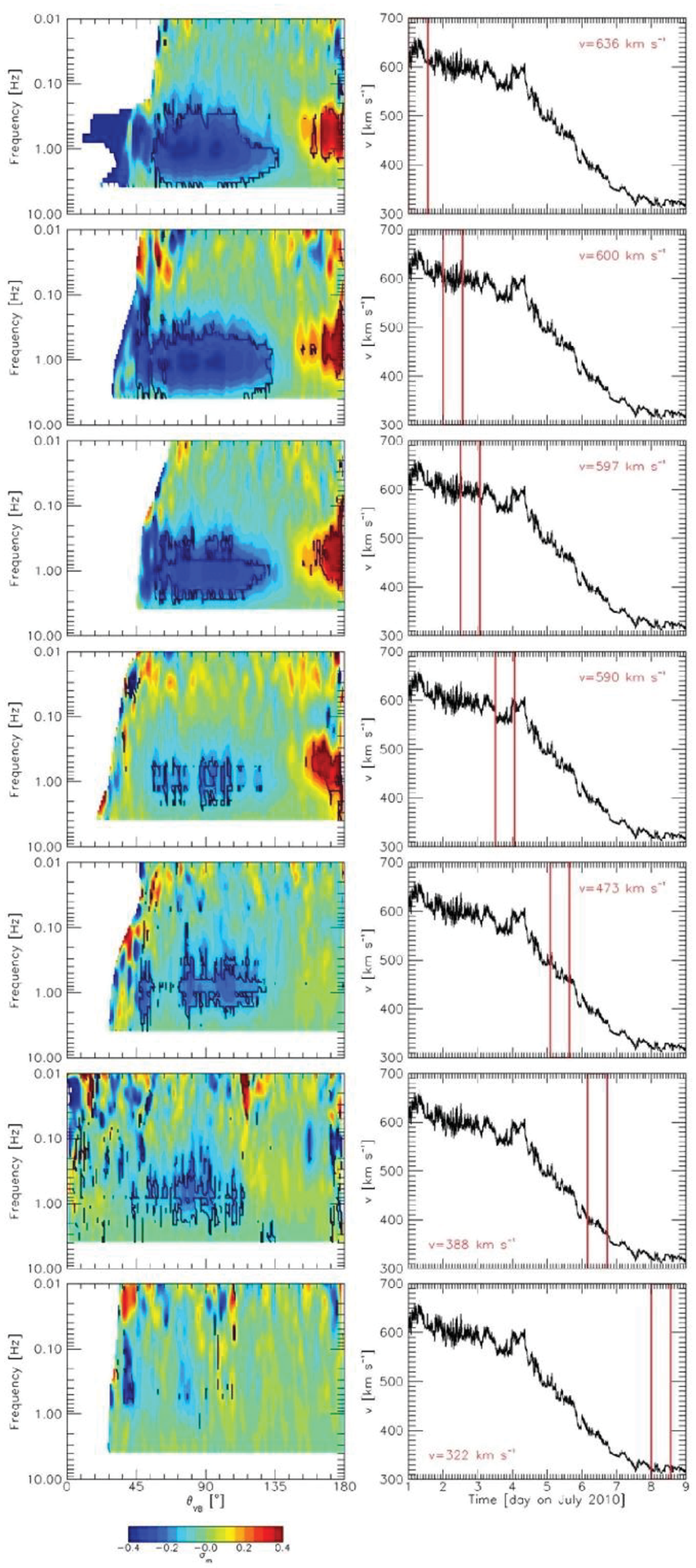}
	\caption{Distributions of the normalized magnetic helicity spectrum with respect to the angle $\theta_{VB}$ between the orientations of the local mean magnetic field and the sampling direction (\emph{left}), observed in the transition from fast to slow wind (\emph{from top to bottom}).
The black contour lines show the 99\% confidence levels. The regions of interest of the analysis along the wind speed profile (\emph{right}) are encircled by red boxes; the average speed of the solar wind in each time interval is also reported.}
	\label{fig:magnetic_helicity}
\end{SCfigure*}

The results of this polarization analysis are shown in Figure \ref{fig:magnetic_helicity}. Each panel of the right-hand-side column shows the speed profile of the same high speed stream and the locations of the analysed $12 \textrm{hr}$ time intervals (see Table \ref{tab:data_intervals} for a detailed list of intervals). The distributions of $\sigma_{m}(\theta_{VB},\tau)$ are shown in the corresponding panels on the left-hand-side column.

\begin{table}
	\caption{Starting time and average solar wind speed $V_{sw}$ of the data intervals used in the present analysis. Each interval lasts 12 hr}
	\begin{center}
		\begin{tabular}{cccc}
			\hline
			Day of July 2010 & DoY & Time & $V_{sw}$ \\
			                        & [day]    & [hh:mm] & [$km\,s^{-1}$] \\
			\hline
			01 & 182 &00:00 & 636 \\
			02 & 183 &00:00 & 600 \\
			02 & 183 &12:00 & 597 \\
			03 & 184 &12:00 & 590 \\
			05 & 186 &02:00 & 473 \\
			06 & 187 &04:00 & 388 \\
			08 & 189 &00:00 & 322 \\
			\hline
		\end{tabular}
		\label{tab:data_intervals}
	\end{center}
\end{table}

The first time interval shows a clear right-handed signature around  $90^\circ$ and a less extended left-handed signature around  $180^\circ$.
The $99\%$ confidence level is indicated by the solid black contour lines. The probability that the results encircled by these contours might be obtained by pure chance is only $1\%$ \citep[see][for a detailed discussion]{telloni2015}.

The color saturation indicates that these two populations are strongly polarized.
As already reported in literature \citep{he2011,podesta2011,he2012a,he2012b,telloni2015}, the right- and left-handed polarized magnetic fluctuations sampled quasi-perpendicularly and quasi-antiparallely to the local magnetic field direction, should be associated to KAWs and Alfv\'{e}n-ioncyclotron waves, respectively .

This polarization persists throughout the trailing edge of the stream, where the Alfv\'{e}nic correlation shown in Figure \ref{fig:correlation_factor} is higher, although some decrease in the extension and in the normalized magnetic helicity intensity can be noticed as we move along the speed profile.
In particular, between the $4th$ and the $5th$ intervals, i.e. when we enter the rarefaction region of the stream, the left-handed polarization is lost and the right-handed one is much reduced. As we move towards lower and lower speed we start loosing also the right-handed polarized population to the extent that in the last interval (bottom panel) our technique does not reveal the presence of any kind of polarized fluctuations.

However, we agree with \citet{he2011} about a possible bias of reduced magnetic helicity results in favor of all those fluctuations propagating quasi along the radial direction, i.e. the sampling direction. This needs to be accounted for in future studies estimating the relative contribution to the reduced magnetic helicity of those fluctuations propagating in different directions, i.e. at a given angle with the sampling direction.



This same technique allows us to estimate the power associated to the highly oblique and quasi-antiparallel polarized fluctuations with respect to the local mean field direction.
To do so, we averaged the total power spectrum of the magnetic fluctuations (not shown) in angular intervals
around $\theta_{VB}\sim90^{\circ}$ and close to $\theta_{VB}180^{\circ}$, respectively, i.e. within the intervals [$70-110^{\circ}$] and [$140-180^{\circ}$],
in order to include most of the two populations.
For the quasi parallel interval we could not center the $180^\circ$ value in the middle due to the fact that, in our case, there are no estimates available at small angles with respect to the magnetic field direction.

\begin{figure*}
	\centering
	\includegraphics[width=\hsize]{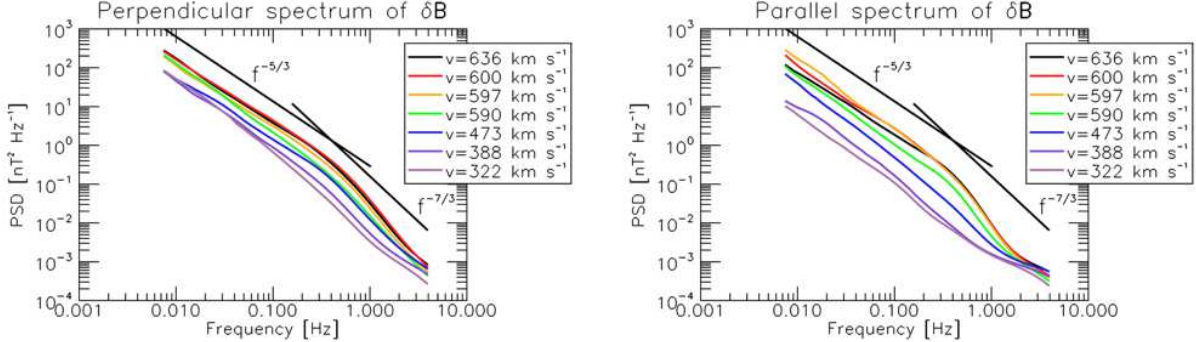}
	\caption{Power Spectral Density (PSD) of perpendicular (\emph{left-hand-side panel}) and parallel (\emph{right-hand-side panel}) magnetic fluctuations as inferred along the high speed stream; different colors are used for different solar wind samples characterized by different average speeds as reported in both panels. A scaling of $-5/3$ \citep[typical of the Kolmogorov turbulence,][]{kolmogorov1941} and of $-7/3$ \citep[expected for the incompressible Hall effect,][]{galtier2007} for the fluid and kinetic ranges, respectively, are shown for reference in both panels.}
	\label{fig:psd}
\end{figure*}

The left and right panels of Figure \ref{fig:psd} show the Power Spectral Density (PSD) for the perpendicular and parallel fluctuations, respectively, within the various time intervals  corresponding to the solar wind samples shown in the right-hand-side columns of Figure \ref{fig:magnetic_helicity}. The different time intervals are indicated by different colors.

The power density spectra for both perpendicular and parallel fluctuations, $P_\perp$ and $P_\parallel$, respectively, are generally higher within the fast trailing edge of the stream if compared to the slower rarefaction region. As a matter of fact, the first four intervals show the highest spectral density.
Within the inertial range, parallel and perpendicular fluctuations share approximately the typical $-5/3$ Kolmogorov scaling \citep{kolmogorov1941}, regardless of the analyzed wind sample, either fast or slow but, the perpendicular power is generally larger,
as predicted by \cite{goldreich1995} for anisotropic turbulence, although a more direct comparison would require narrower angular bins like in \citet{horbury2008}.

At frequencies beyond the spectral break  located between at $0.3-0.4\,Hz$, $P_\parallel$ is generally steeper than $P_\perp$. In addition, as already reported in literature \citep{bruno2014b}, a large variability of the spectral index is observed at proton scales. The value of the spectral index depends on the power associated with the fluctuations within the inertial range: the higher the power, the steeper the slope.

\begin{figure*}
	\centering
    \includegraphics[scale=1.0]{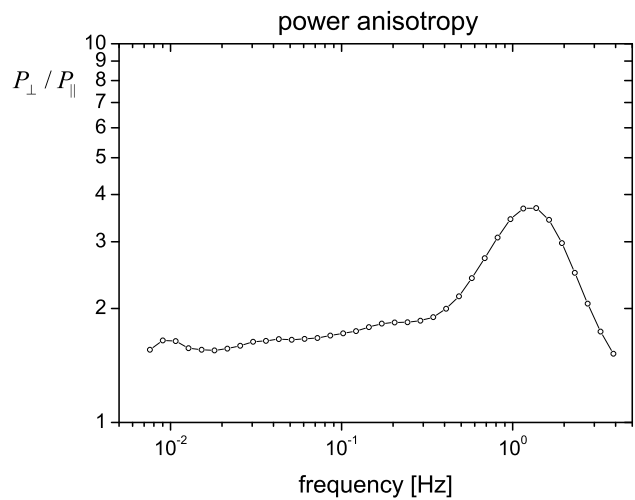}
	\caption{Anisotropy ratio between perpendicular $P_\perp$ and parallel $P_{//}$ power density. See text for details.}
	\label{fig:aniso}
\end{figure*}

In Figure \ref{fig:aniso} we report the ratio between perpendicular $P_\perp$  and parallel $P_{//}$ power density, having averaged together the first four spectra within the high speed trailing edge. This ratio shows an anisotropy in favor of $P_\perp$ throughout the frequency range. In particular, this anisotropy increases for increasing frequency and, around $1$Hz reaches its maximum value before decreasing dramatically right after. These results are qualitatively similar to those obtained by \citet{podesta2009}. In our case, the maximum value of this ratio coincides with the frequency location of the core of the KAW population which is at slightly higher frequency with respect to the core of the parallel population.

\begin{figure*}
	\centering
    \includegraphics[width=\hsize]{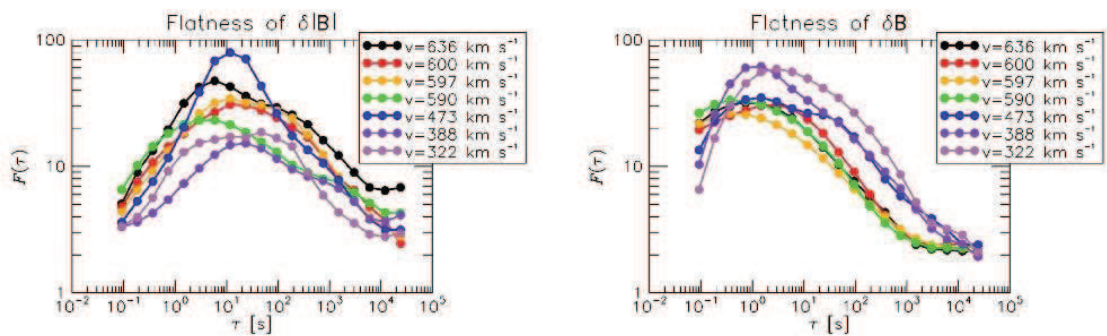}
	\caption{left-hand-side panel: flatness factor versus time scale $\tau$, relative to fluctuations of the magnetic field intensity observed within the time intervals reported in Table \ref{tab:data_intervals}; the legend reports the average speed and the corresponding color for each time interval. Right-hand-side panel: flatness factor relative to fluctuations of the magnetic field vector in the same format as of the left-hand-side panel}
	\label{fig:intermittency}
\end{figure*}

Finally, Figure \ref{fig:intermittency} reports the results relative to the intermittency analysis of magnetic field intensity and vector fluctuations for the same time intervals listed in Table \ref{tab:data_intervals}.
For details on the methodology based on the flatness and for a review on previous intermittency results related to magnetic field fluctuations within both fast and slow wind, the reader can refer to the paper by \cite{bruno2003}.

For each time interval, we show values of the flatness of the distributions of the magnetic field intensity and vector differences versus time scale.

In the left-hand-side panel, flatness increases for all intervals when we move from large to small scales, reaching its maximum at scales slightly larger than the inverse of the frequency break which, in this case, is around 0.3-0.4 Hz \citep{bruno2014a}. Beyond this frequency, all the curves experience a rapid decrease towards the Gaussian value of the flatness. There is no much order in the way the curves are organized in the plot. Following the definition of intermittency given by \cite{frisch1995}, i.e. the same time series is defined to be more intermittent if the flatness grows faster with decreasing the scale, it is rather difficult to establish in a reliable way which curve is more intermittent.

On the contrary, the same parameter computed for directional fluctuations, shown in the right-hand-side panel, shows that the flatness of slow wind starts to increase before that of fast wind confirming the more intermittent nature of its magnetic fluctuations \citep{marsch1993, bruno2003}. Moreover, again at odds with compressive fluctuations, the decrease of these curves seems to take place at scales slightly smaller, closer to the inverse of the frequency break.

\section{Discussion and conclusions}
\label{sec:discussion_conclusions}

The character of the fluctuations within roughly the first decade of frequency beyond the high frequency break separating fluid and  kinetic regimes, strongly depends on the wind type. Our analysis indicates that both wind speed and Alfv\'{e}nicity are valid discriminants as we found remarkable differences between fast and slow wind.
We find a clear signature of both KAW and Ion-Cyclotron waves within those regions of the trailing edge characterized, at fluid scales, by large fluctuations with a strong Alfv\'{e}nic character. As a matter of fact, the first four intervals, characterized by similar Alfv\'{e}nicity and similar amplitude of the fluctuations, show rather similar fluctuations at kinetic scales but, as soon as the corresponding Alfv\'{e}nic correlation $C_{VB}$ and the fluctuations' amplitude decrease, the polarization signature of these fluctuations is clearly depleted.  Moreover, as we enter the rarefaction region and move towards slower regions, not only the ion-cyclotron helicity signature disappears followed by a more gradual disappearance of the KAW  but, parallel and perpendicular spectral analyses reveal that also the associated power strongly decreases. In addition, the spectral analysis has not shown meaningful differences between the spectral indices related to parallel and perpendicular fluctuations within the inertial range since both classes of fluctuations seem to follow roughly a $-5/3$ scaling.
Following the critical balance predictions \citep{goldreich1995} we would expect a scaling of $-2$ for parallel fluctuations and $-5/3$ for the perpendicular ones
but, our time intervals are probably too short and the angular bins too wide to allow for
this kind of comparison. Nevertheless, there is no doubt that there is a clear anisotropy in favor of the perpendicular spectrum and, at least within the kinetic regime, the parallel spectrum is the one the experiences the largest decrease as also shown by the faster disappearance of the left-handed fluctuations.
One possible mechanism for the disappearance of the fluctuations with $k_\|$ might be the ion-cyclotron resonance which was recently re-invoked as a possible dissipation mechanism by  \citet{bruno2014a}[see other references therein]. On the other hand, the depletion of KAW seems to be less dramatic since their polarization signature tends to survive in a large fraction of the rarefaction region.
The intermittency observed within the various time intervals is consistent with results already reported in literature \citep{wu2013,telloni2015}. For each of the analysed time intervals, flatness increases from large to small scales down to scales corresponding roughly to the high frequency break (see discussion in the previous section). However, around and beyond the frequency break, the flatness starts to decrease. In particular, we noticed that while compressive fluctuations become less intermittent quite before the frequency break, as already reported by \citep{wu2013}, directional fluctuations have a much slower decrease which starts around the frequency break. We believe that there might be some connection between this observation and the fact that the central part of the KAW and Alfv\'{e}n-ion-cyclotron populations is generally located at scales slightly smaller than the one corresponding to the spectral break as reported in Figure 2 from \cite{telloni2015}. However, this conclusion needs to be corroborated by further investigations since it seems not to be supported by the fact that whilst in slow wind the signatures of both KAW and ion-cyclotron waves gradually disappear  we still register a roughly similar behavior of intermittency. Probably, within slow wind there is still some residual population of KAW and ion-cyclotron waves that our analysis is not able to unravel because of the low level of the corresponding signals.

\begin{acknowledgements}
This research was partially supported by the Italian Space Agency (ASI) under contracts I/013/12/0 and I/022/10/2, and by the European Commission's Seventh Framework Program under the grant agreement STORM (project n$^{\circ}$ 313038). Data from WIND were obtained from NASA-CDAWeb website.
\end{acknowledgements}


\begin{thebibliography}{}
	\bibitem[Alexandrova et al., 2008]{alexandrova2008}
Alexandrova, O., Carbone, V., Veltri, P., \& Sorriso-Valvo, L. 2008, \apj, 674, 1153
	\bibitem[Alexandrova et al., 2013]{alexandrova2013}
Alexandrova, O., Chen, C. H. K., Sorriso-Valvo, L., Horbury, T. S., \& Bale, S. D. 2013, \ssr, 178, 101
    \bibitem[Bavassano et al.(1998)]{bavassano1998}
Bavassano, B., Pietropaolo, E., Bruno, R.\ 1998.\jgr, 103, 6521.
	\bibitem[Bruno et al., 2003]{bruno2003}
Bruno, R., Carbone, V., Sorriso-Valvo, L., \& Bavassano, B. 2003, \jgr, 108, 1130
	\bibitem[Bruno et al., 2008]{bruno2008}
Bruno, R., Pietropaolo, E., Servidio, S., et al. 2008, AGU Fall Meeting 2008, SH42A-06
	\bibitem[Bruno \& Carbone, 2013]{brunocarbone2013}
Bruno, R., \& Carbone, V. 2013, Living Rev. Solar Phys., 10, 2
	\bibitem[Bruno \& Trenchi, 2014]{bruno2014a}
Bruno, R., \& Trenchi, L. 2014, \apjl, 787, L24
	\bibitem[Bruno et al., 2014]{bruno2014b}
Bruno, R., Trenchi, L., \& Telloni, D. 2014, \apjl, 793, L15
	\bibitem[D'Amicis \& Bruno, 2015]{damicis2015}
D'Amicis, R., \& Bruno, R. 2015, \apj, 805, 84
	\bibitem[Denskat et al., 1983]{denskat1983}
Denskat, K. U., Beinroth, H. J., \& Neubauer, F. M. 1983, \jgr, 54, 60
	\bibitem[Frisch, 1995]{frisch1995}
Frisch, U. 1995, \emph{Turbulence: The Legacy of A. N. Kolmogorov}, Cambridge University Press, New York
    \bibitem[Galtier and Buchlin(2007)]{galtier2007}
Galtier, S., Buchlin, E., 2007, \apj, 656, 560-566.
	\bibitem[Gary \& Borovsky, 2004]{gary2004}
Gary, S. P., \& Borovsky, J. E. 2004, \jgr, 109, 6105
	\bibitem[Gary \& Smith, 2009]{gary2009}
Gary, S. P., \& Smith, C. W. 2009, \jgr, 114, A12105
	\bibitem[Goldreich \& Sridhar, 1995]{goldreich1995}
Goldreich, P., \& Sridhar, S. 1995, \apj, 438, 763
	\bibitem[Goldstein et al., 1994]{goldstein1994}
Goldstein, M. L., Roberts, D. A., \& Fitch, C. A. 1994, \jgr, 99, 11519
	\bibitem[Hamilton et al., 2008]{hamilton2008}
Hamilton, K., Smith, C. W., Vasquez, B. J., \& Leamon, R. J. 2008, \jgr, 113, A01106
	\bibitem[He et al., 2011]{he2011}
He, J., Marsch, E., Tu, C., Yao, S., \& Tian, H. 2011, \apj, 731, 85
	\bibitem[He et al., 2012a]{he2012a}
He, J., Tu, C., Marsch, E., \& Yao, S. 2012, \apj, 749, 86
	\bibitem[He et al., 2012b]{he2012b}
He, J., Tu, C., Marsch, E., \& Yao, S. 2012, \apjl, 745, L8
	\bibitem[Horbury et al., 2008]{horbury2008}
Horbury, T. S., Forman, M., \& Oughton, S. 2008, \prl, 101, 175005
    \bibitem[Hundhausen(1972)]{hundhausen1972}
Hundhausen, A.~J.\ 1972.\ Physics and Chemistry in Space 5
	\bibitem[Karimabadi et al., 2013]{karimabadi2013}
Karimabadi, H., Roytershteyn, V., Wan, M., et al. 2013, Physics of Plasmas, 20, 012303
	\bibitem[Kiyani et al., 2009]{kiyani2009}
Kiyani, K. H., Chapman, S. C., Khotyaintsev, Y. V., Dunlop, M. W., \& Sahraoui, F. 2009, \prl, 103, 075006
	\bibitem[Kiyani et al., 2013]{kiyani2013}
Kiyani, K. H., Chapman, S. C., Sahraoui, F., et al. 2013, \apj, 763, 10
	\bibitem[Kolmogorov, 1941]{kolmogorov1941}
Kolmogorov, A. N. 1941, Dokl. Akad. Nauk. SSSR, 30, 301
	\bibitem[Leamon et al., 1998]{leamon1998}
Leamon, R. J., Smith, C. W., Ness, N. F., Matthaeus, W. H., \& Wong, H. K. 1998, \jgr, 103, 4775
	\bibitem[Lepping et al., 1995]{lepping1995}
Lepping, R. P., Ac$\tilde{\textrm{u}}$na, M. H., Burlaga, L. F., et al. 1995, \ssr, 71, 207
	\bibitem[Marsch et al., 1982]{marsch1982}
Marsch, E., Schwenn, R., Rosenbauer, H., et al. 1982, \jgr, 87, 52
	\bibitem[Marsch, 2006]{marsch2006}
Marsch, E. 2006, Living Rev. Solar Phys., 3, 1
	\bibitem[Marsch, 2012]{marsch2012}
Marsch, E. 2012, \ssr, 172, 23
    \bibitem[Marsch and Liu, 1993]{marsch1993}
Marsch, E. and Liu, S., 1993, Ann. Geophys., 11, 227–238.
	\bibitem[Matthaeus \& Goldstein, 1982]{matthaeus1982}
Matthaeus, W. H., \& Goldstein, M. L. 1982, \jgr, 87, 6011
	\bibitem[Matteini et al., 2014]{matteini2014}
Matteini, L., Horbury, T. S., Neugebauer, M., \& Goldstein, B. E. 2014, \grl, 41, 259
    \bibitem[Ogilvie et al.(1995)]{ogilvie1995}
Ogilvie, K.~W., et al., \ssr, 71, 55
    \bibitem[Narita et al.(2009)]{narita2009}
Narita, Y., Kleindienst, G., Glassmeier, K.-H.\ 2009.\ Annales Geophysicae 27, 3967-3976.
	\bibitem[Perri et al., 2012]{perri2012}
Perri, S., Goldstein, M. L., Dorelli, J. C., \& Sahraoui, F. 2012, \prl 109, 191101
	\bibitem[Podesta \& Gary, 2011]{podesta2011}
Podesta, J. J., \& Gary, S. P. 2011, \apj, 734, 15
    \bibitem[Podesta, 2009]{podesta2009}
Podesta, J.~J.\ 2009, \apj, 698, 986
	\bibitem[Salem et al., 2012]{salem2012}
Salem, C. S., Howes, G. G., Sundkvist, D., et al. 2012, \apjl, 745, L9
	\bibitem[Servidio et al., 2012]{servidio2012}
Servidio, S., Valentini, F., Califano, F., \& Veltri, P. 2012, \prl, 108, 045001
	\bibitem[Sahraoui et al., 2009]{sahraoui2009}
Sahraoui, F., Goldstein, M. L., Robert, P., \& Khotyaintsev, Y. V. 2009, \prl, 102, 231102
	\bibitem[Telloni et al., 2015]{telloni2015}
Telloni, D., Bruno, R., \& Trenchi, L. 2015, \apj, in press
	\bibitem[TenBarge et al., 2012]{tenbarge2012}
TenBarge, J. M., Podesta, J. J., Klein, K. G., \& Howes, G. G. 2012, \apj, 753, 107
	\bibitem[Torrence \& Compo, 1998]{torrence1998}
Torrence, C., \& Compo, G. P. 1998, Bull. Am. Meteorol. Soc., 79, 61
    \bibitem[Tu and Marsch(1995)]{tumarsch1995}
Tu, C.-Y., Marsch, E.\ 1995 \ssr 73, 1
	\bibitem[Turner et al., 2011]{turner2011}
Turner, A. J., Gogoberidze, G., Chapman, S. C., Hnat, B., \& M$\ddot{\textrm{u}}$ller, W.-C. 2011, \prl, 107, 095002
	\bibitem[Wan et al., 2012]{wan2012}
Wan, M., Matthaeus, W. H., Karimabadi, H., et al. 2012, \prl, 109, 195001
	\bibitem[Wu et al., 2013]{wu2013}
Wu, P., Perri, S., Osman, K., et al. 2013, \apjl, 763, L30
\end{thebibliography}
\end{document}